\documentclass[aps,pre,twocolumn,noshowpacs,noshowkeys,square,numbers,amssymb,amsmath,nobibnotes,footinbib]{revtex4}
\usepackage[utf8]{inputenc}%
\usepackage[T1]{fontenc}%
\usepackage{bm}%
\usepackage[colorlinks=true,linkcolor=red]{hyperref}
\usepackage{amsmath}
\usepackage{amsfonts}
\usepackage{amssymb}
\usepackage{verbatim}
\usepackage{epsfig}
\usepackage{eucal}
\usepackage{graphicx}
\usepackage{soul}
\usepackage[sort&compress]{natbib}
\usepackage{times}
\usepackage{lipsum}
\usepackage{mathtools}
\usepackage[dvipsnames]{xcolor}

\begin{document}


\title{Maximal modularity and the optimal size of parliaments}


\author{Luca Gamberi}
\author{Yanik-Pascal F\"orster}
\author{Evan Tzanis}
\author{Alessia Annibale}
\author{Pierpaolo Vivo}

\affiliation{Department of Mathematics and Quantitative and Digital Law Lab, King’s College London, London WC2R 2LS, United Kingdom}



\date{\today}

\begin{abstract}
An important question in representative democracies is how to determine the optimal parliament size of a given country. According to an old conjecture, known as the cubic root law, there is a fairly universal power-law relation, with an exponent close to 1/3, between the size of an elected parliament and the country's population. Empirical data in modern European countries support such universality but are consistent with a larger exponent. In this work, we analyze this intriguing regularity using tools from complex networks theory. We model the population of a democratic country as a random network, drawn from a growth model, where each node is assigned a constituency membership sampled from an available set of size $D$. We calculate analytically the modularity of the population and find that its functional relation with the number of constituencies is strongly non-monotonic, exhibiting a maximum that depends on the population size. The criterion of maximal modularity allows us to predict that the number of representatives should scale as a power-law in the size of the population, a finding that is qualitatively confirmed by the empirical analysis of real-world data.
\end{abstract}


\maketitle


\section{Introduction}

In modern times, representative democracies have played a leading role in the advancement of human rights, education, and technology on a global scale.  At the heart of every representative democracy is a centralized parliament: an assembly of elected citizens who are delegated by their constituents to exercise the legislative power, and to keep the government in check \cite{Rush-book}. This apparatus has an operating cost and, in the shadows of political scandals, economic crises, and social turmoil, people have questioned the effectiveness of their country’s costly political and administrative structure and have claimed that a reduction of the number of elected representatives would reduce deviant behaviors and enhance efficiency of parliamentary works \cite{L-B12}. However, there is so far no sound analytical framework to determine the optimal parliament size of a given country, so to ensure an adequate representation and cost-effectiveness, which are both in the public interest.
In this paper, we argue that a \textit{principle of maximal modularity} can provide some reliable guidance on how to determine the absolute number of representatives required for efficient public representation in a democratic country. This principle may therefore provide a transparent reference point to inform public policies.

Generally speaking, the ideal number of members of Parliament (MPs) has to strike a balance between \textit{efficiency}, in terms of the share of power held by each MPs and their ability to realize their electoral agenda, and \textit{optimal representativity}, i.e. the ability of the MPs to promote the instances of their voters, in proportion to their number. Both criteria are encoded in the assembly size, as a bigger chamber allows constituencies to be smaller and thus more homogeneous in terms of character, local economic activity, and social needs. On the other hand, it diminishes the influence and resources that each MP can count on to advance their agenda and thus promote their constituents' interests \cite{RW-book}. 
The ``efficiency'' paradigm has been at the core of a flourishing line of research amongst political scientists and “electoral engineers”, since the late 80s \cite{Farr-book}.  Researchers have revealed the effect of different electoral systems on the efficiency and stability of political architecture, in relation to the size of the corresponding assembly \cite{TS89}. Representation of minority groups, gender quotas, ballot votes, and district sizes are believed to heavily influence the efficiency of a parliament and the relative voting power of political parties \cite{JO15,Lij06}.
Another pressing issue that has been thoroughly studied concerns the distribution of relative weights of votes for delegates in international bodies or for parliaments in federal states such as the United States. The most famous approach is the one proposed  by Webster and Sainte-Laguë independently, after which several other quotient rules were adopted \cite{McL08}. Game theory approaches, such as the Penrose square root law \cite{Pen46}, were also proposed later on and are currently used for instance in the Council of the European Union, to implement a “one person, one vote” system \cite{SZ06}. 
An interesting empirical decision-making model linking participation in elections and electoral college size, at any level (local to national), was proposed in \cite{BHL13}.

Both the problem of efficiency and relative representativity have been investigated for a long time in the political science literature and share a common denominator: they depend - directly or indirectly - on the absolute chamber size in a way that is yet to be fully understood \cite{JO15}.

\begin{figure}[htb]
	\centerline{\epsfig{figure=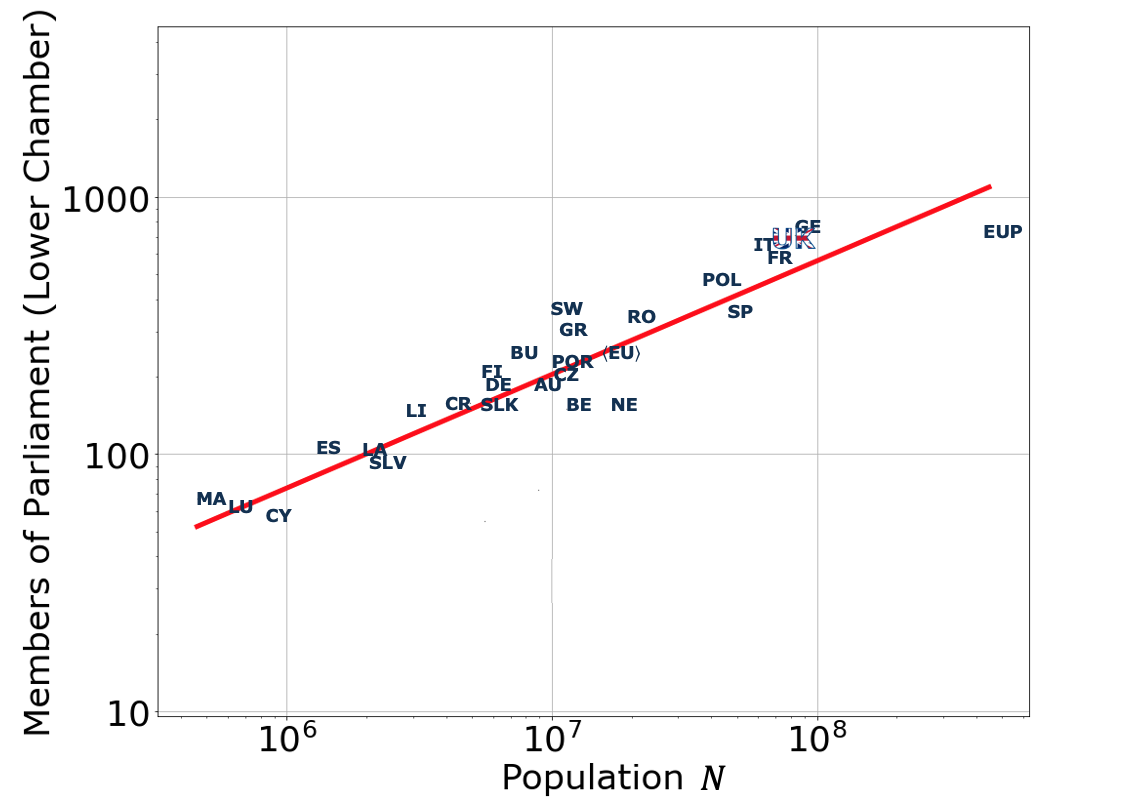 ,width=0.45\textwidth}} \caption{The figure shows a log-log scale plot of the size of lower chamber $S$ vs. population $N$ for some European countries and the EU Parliament. The best fit with a power-law (red line) shows that the size of the chamber grows as $S\approx \alpha N^{\gamma}$, with $\gamma \approx 0.44$ and $\alpha \approx 0.17$. Demographic data from Eurostat (2017).\label{fig:euparl_vs_pop}}
\end{figure}

In recent times, political scientists and technocrats have heavily relied on the so-called \emph{cubic root law} (CRL) formulated by Taagepera and collaborators in \cite{Taa72,TaSh-book,TR02,Taa-book}. This law follows the realization that the size of most elected parliaments exhibits a strong statistical regularity with respect to their population. The proposed empirical model optimizes the assemblies' representation based on the efficiency of communication between MPs and their constituents.
According to Taagepera's arguments, the size of parliaments should follow $S\propto N_0^{\gamma}$, where $\gamma = 1/3$ and $N_0$  an “effective” population size, rescaled by considering only the portion of active voters and, among them, the fraction of literate adults \cite{Taa72}. As literacy is believed to be strongly correlated with mobility, the latter rescaling was introduced to account for social mobility in the absence of a reliable direct measure for this parameter \cite{Russ-book}. Figure \ref{fig:euparl_vs_pop} highlights the aforementioned regularity, but when considering the sizes of European lower chambers only, the best fitting curve deviates from the theoretical CRL, resulting in $\gamma \approx 0.44$. A more comprehensive analysis of parliaments' size data can be found in \cite{BHL13}. It is also worth mentioning that the CRL formulated in terms of an effective population was perceived to be in contrast with the spirit of any “good” representation model that should include the entire pool of constituents, regardless of age, political engagement, or education \cite{JO15}.

In this work, we tackle the democratic representation problem using network theory. Networks have been successfully employed in social science for over 30 years, for their versatility in describing different aspects of political, behavioral, and social interaction between individuals \cite{NBW-book}. In social networks, nodes represent social agents (for example, individuals in a population) and links represent their interactions. The network structure contains important information about relational ties in a society and is likely influenced by agents' attributes (e.g. age, occupation, wealth ...) \cite{Wass-book}. An important topological feature of real-world social networks is their \textit{scale-free} degree distribution, i.e. the distribution of the number of ties following a power-law \cite{Cald-book}. This topology can be reproduced in growing networks using a preferential attachment wiring protocol that was proposed in \cite{BA99}. 

The objective of this work is to shed light on the observed statistical regularities in the size of parliaments. We will be focusing on electoral systems in which representatives are elected according to an FPTP (``First-Past-The-Post'') principle, i.e. whoever collects the majority of votes within a constituency gets elected. However, we expect our results to be more broadly applicable.

Taking the United Kingdom as an example, each Member of Parliament is elected to the House of Commons from one of the 650 constituencies. The nature and physical boundaries of the constituencies are regulated by the \textit{House of Commons (Redistribution of Seats) Act} (1944), which prescribes that the MP’s role is to ``represent the common interest of the residents in a spatially bounded territory''. Thus, when constituencies are designed, the legislator should aim to enclose within geographical boundaries areas that share common interests and values \cite{RJP13}.  

The network model we propose is inspired by this design principle. We build a synthetic scale-free network in which $N$ agents (nodes) represent the entire population of a country that has to be partitioned into constituencies, each electing their MP to the national Parliament. Two citizens are connected if there is a stable social interaction between them. Individuals are therefore arranged into social communities, as proposed, for instance, in \cite{WBB76}. 
The key result of our paper is a principle for determining the optimal number of constituencies, i.e. for grouping the population into electoral clusters that ``best'' represent the underlying community structure of the network. 
We remark that we need to strike a balance between representativity and homogeneity in constituency size. Hence, our approach needs to improve upon the standard ``community detection'' framework, which would allow the constituency size to fluctuate wildly. We achieve this result by constraining the size of the constituencies at the outset, and determining the number of constituencies that optimizes the partitioning of the underlying network.

More specifically, we generate synthetic networks from a growth model with preferential attachment to nodes of higher degrees and within the same constituency. We introduce a \textit{mobility} (or affinity) parameter into our model, of a similar nature to that proposed in \cite{Taa72}, which allows us to tune the probability that a node interacts with foreign constituencies. Note that in this approach, each node is assigned a constituency label a priori and the network topology follows as a result of this assignment, in contrast to what usually happens in community detection, where node memberships are determined a posteriori, based on the network topology. In this regard, our approach is based on a generative model for network clustering rather than a discriminative model.

Working with synthetic networks relieves us from making assumptions over geographical constraints, as constituencies are purely virtual, i.e. designed around groups of people with stronger interpersonal ties. Although this modeling choice may need to be supplemented with more realistic assumptions, non-geographical electoral systems have been proposed in the past with strong supporting arguments in terms of representation of minorities and dispersed communities \cite{Rahf-book}. Furthermore, geographical constraints would strongly depend on the country at hand, whereas our model aims to be as general as possible. 

We adopt the \emph{modularity} as a metric to measure the \textit{goodness} of these partitions and we derive an exact expression for the average network modularity, in terms of the number $D$ of constituencies, for fixed network size $N$. By maximizing the modularity, we are able to determine analytically the optimal number of equally sized constituencies into which networks generated according to our prescription should be partitioned.
In our investigation, we find that the empirical regularities discussed above arise quite naturally from the topology of the clustered networks that we study here. 

The manuscript is organized as follows: in Section II we introduce the network growth model. In Section II(A) we derive and solve the recursive equation for the expected modularity at generic network size and number of constituencies. In Section II(B) we present a numerical solution for the maximum modularity as a function of the network size and we construct an approximate scheme to solve the problem analytically. Finally, we present our main findings in the Conclusion and we compare with empirical evidence. The technical details of our derivation are presented in the Appendix.

Our findings reveal that the optimal partitioning in constituencies for a given population is well approximated by a power-law $S\simeq N^{\gamma}$, which is in qualitative agreement with the empirical data. Interestingly, we observe that the mobility does not play a significant role in determining the exponent $\gamma$, at least for the homogeneous mobility case studied in this work. 

\section{The model}
We model social interactions within a population by means of simple, undirected networks, which are constructed using a modified version of the Barab{\'a}si-Albert (BA) algorithm \cite{BA99,DM02}. 

\begin{figure}[h]	
	\centering
	{\includegraphics[width=0.48\textwidth]{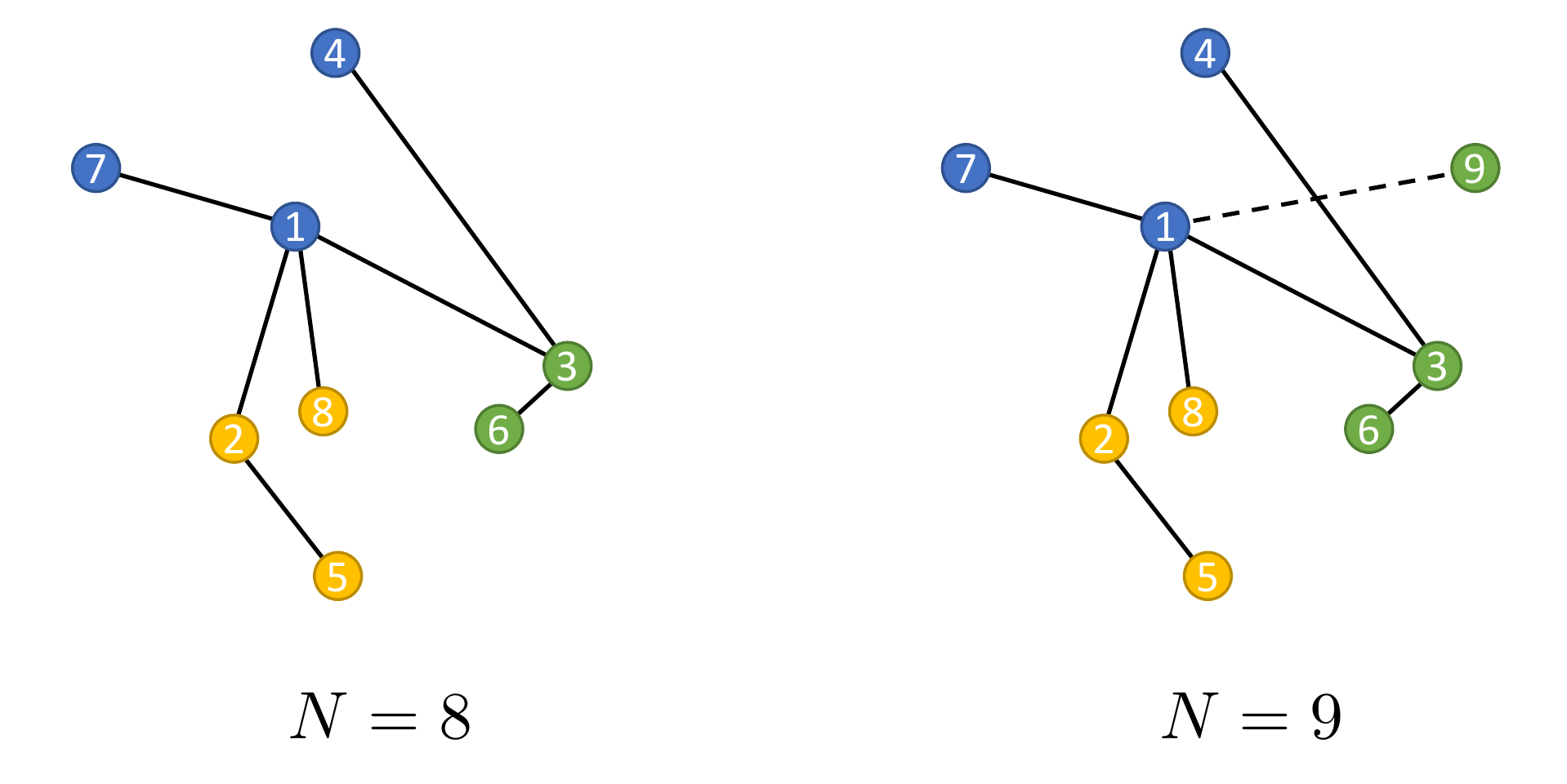}}
	\caption{Sketch of a $D=3$ constituencies network, for the case $m=1$, at the steps $N=8$ (left) and $N=9$ (right) of the growth algorithm, respectively. Membership attributes 1, 2 and 3  represented by the colour blue, yellow and green respectively are allocated sequentially in such a way that $\sigma_i=\mod(1+i, 3)$, hence $\sigma_1=1$, $\sigma_2=2$, $\sigma_3=3$, $\sigma_4=1$ and so on. The dashed line represents the new connection made at step $N=9$.}
	\label{fig:net_growth}
\end{figure}

The network is formed dynamically in such a way that at each time step $N$ a node $N$ is created, with $m$ stubs, and a constituency membership label $\sigma_N\in\{1,...,D\}$ is assigned, according to a prescribed sequential order such that $\sigma_N=\textrm{mod}(1+N, D)$ (see Figure \ref{fig:net_growth} for an illustration). In this way, any two nodes $i$ and $i+D$ have the same membership and all the constituencies have roughly equal size (their sizes are either identical or differ by one unit). The sequential assignment is a modeling choice that greatly simplifies the analytical treatment presented in this section, however, the final outcome does not heavily depend on the way the constituencies $\sigma$'s are assigned, provided that they are on average all equally sized. The network at time step $N$ is represented by an $N\times N$ adjacency matrix $\bm{A}(N)$, with entries $A_{ij}(N)$ for $i,j\leq N$. 

When a node $N$ is added, each of the $m$ stubs is wired to a random node $i$ of the existing network sampled with probability
\begin{equation}
	p_{iN}=m\frac{\sum_{\ell=1}^{N-1} A_{i,\ell}(N-1)}{L(N-1)/D}p(\sigma_i|\sigma_N)\ ,
	\label{eq:mobility_attachment_prob}
\end{equation}
with $L(N) = \sum_{i=1}^n k_i(N)  = 2m(N - m)$ being the total number of links present in the network,  $k_i(N) = \sum_{j=1}^N A_{i,j}(N)$ the degree of node $i$, calculated at time $N$, and $p(\sigma_i|\sigma_N)$ being the probability that any node with given constituency label $\sigma_{N}$ attaches to any of the nodes with constituency $\sigma_i$, similarly to a block model \cite{HLL83}. As the addition of new nodes cannot modify the links between pre-existing nodes, we have that $A_{ij}(N)$ is the same for any $N\geq \max(i,j)$, so from now on, we will drop the time index from the entries of the adjacency matrix. We prescribe that the initial configuration of the network growth be a clique of $m+1$ nodes. Accordingly, we set the initial time at $m+1$, so that the growth process starts at $m+2$.

The probability $p(\sigma_i|\sigma_N)$, parametrized by a mobility parameters $\mu$ that controls the likelihood to pick the target constituency, is given by
\begin{equation}
p(\sigma_i|\sigma_N) = \frac{\mu}{D} + (1-\mu)\delta_{\sigma_i,\sigma_N}\ ,
\label{eq:prob_si_sN}
\end{equation}
which is normalized $\sum_{\sigma_i=1}^D p(\sigma_i|\sigma_N) =1$, as it should. 

Hence, for $\mu = 0$, the new node $N$ will attach necessarily to a member of its own community whereas, for $\mu=1$, $N$ can attach to any community with the same probability $1/D$. 
Thus, the probability that a new node attaches to a given node of a foreign constituency is $\mu/D$. The contribution $\sum_{\ell=1}^{N-1} A_{i,\ell} /L (N-1)$ in the definition \eqref{eq:mobility_attachment_prob} ensures that new nodes attach preferentially to nodes with higher degree, which realizes the scale-free degree distribution that is typical of social networks \cite{ACF09}.

Using Eq. \eqref{eq:mobility_attachment_prob}, we can write the probability for the entry $A_{i,N}$, with $i = 1, ... ,N-1$, of the adjacency matrix, given its previous configuration $\bm{A}(N-1)$ and the community membership sequence denoted by $\bm{\sigma}$, as 
\begin{align}
	p(A_{i,N}|\mathbf{A}(N-1),\boldsymbol{\sigma}(N)) = [p_{iN}\delta_{A_{i,N},1} +\nonumber\\
	+ (1-p_{iN}) \delta_{A_{i,N},0} ]\delta_{A_{i,N},A_{N,i}}\delta_{k_{N}(N),m} \ .
	\label{eq:mobility_ensemble_entry}
\end{align}
Assuming that each of the $m$ stubs is wired independently to a randomly drawn node, the joint distribution for the $N$-th row and column is
 \begin{align}
 &p(A_{i,N}\ldots A_{N-1,N} |\mathbf{A}(N-1),\boldsymbol{\sigma}(N)) =  \nonumber\\
 &\quad \prod_{i=1}^{N-1} p(A_{i,N}|\mathbf{A}(N-1),\boldsymbol{\sigma}(N)) \ ,
 \label{eq:mobility_ensemble_row}
 \end{align}
and, by iteration, one can get the full distribution for the configuration $\bm{A}(N)$ of the adjacency matrix
 \begin{align}
&p(\bm{A}(N) |\boldsymbol{\sigma}(N)) =  \prod_{i=1}^{N-1} p(A_{i,N}|\mathbf{A}(N-1),\boldsymbol{\sigma}(N)) \times \nonumber\\
&\times \prod_{i=1}^{N-2} p(A_{i,N-1}|\mathbf{A}(N-2),\boldsymbol{\sigma}(N-1)) \ldots p(\bm{A}(m+1))\ ,
\label{eq:mobility_ensemble}
\end{align}
with $p(\bm{A}(m+1))=\prod_{i<j}^{m+1} \delta_{A_{i,j},1}\delta_{A_{i,j},A_{j,i}}$, determined by the initial configuration of the growth algorithm. 

Our numerical analysis shows that $m$ does not significantly affect the key observable and results, hence we will limit our analytical considerations to the case $m=1$.

The key observable we will monitor in our model is the modularity, introduced in \cite{NG04}, as a quality factor for a partition of a network in communities. The modularity of a graph is defined as 
\begin{align}
Q_D(N) &=  \frac{1}{L(N)} \sum_{s,r}^N \left[A_{r,s} - \frac{k_r(N)k_s(N)}{L(N)}\right]\delta_{\sigma_r,\sigma_s}\ .
\label{eq:modularity_Newm_def_non_exp}
\end{align}
This quantity compares the intra-cluster edge density of a given network (in our case, the clusters are defined by the constituency membership attribute) with the edge density of a null model, i.e. a set of unbiased random graphs that are wired regardless of the community structure but with the same degree sequence as the original network \cite{New06}. This comparison mechanism provides a reliable metric to establish the goodness of a network clustering procedure. Moreover, the modularity takes values $Q_D(N)\in [-1,1]$, with positive values denoting that a graph exhibits a community structure that is being captured by their assigned memberships \cite{Fort10}.

We will use the modularity to assess the cluster structure induced by the sequence $\bm{\sigma}$ and the underlying social structure originated from the web of connections. We aim to find the number of constituencies that maximizes this observable, resulting in the optimal partitioning of the synthetic population created by the growth algorithm. For a network of size $N$, we have that the expected modularity is given by
\begin{align}
\left<Q_D(N)\right> &=  \frac{1}{L(N)} \left<\sum_{s,r}^N \left[A_{r,s} - \frac{k_r(N)k_s(N)}{L(N)}\right]\delta_{\sigma_r,\sigma_s}\right>\nonumber\\
&=: \frac{1}{L(N)} a_N - \frac{1}{L(N)^2} b_N\ ,
\label{eq:modularity_Newm_def}
\end{align}
where the expectation is over the distribution \eqref{eq:mobility_ensemble}. At the $(N+1)$-th step, one row and one column are added to the adjacency matrix as follows
$$
\bm{A}(N+1) =
\left[\begin{array}{ccccc|c}
&&&&&0\\ &&&&&0\\ &&\bm{A}(N)&&&\vdots\\&&&&&1\\&&&&&0\\\hline 0&0&\cdots &1&0&0\\ \end{array}\right]\ .
$$
We note that the number of links is deterministic as at each time step $m(=1)$ links are added to the network. Therefore we argue that an expression for $\left<Q_D (N+1) \right>$ can be found recursively, in particular by solving recursions for the coefficients $a_N$ and $b_N$ that we present in the following subsection.

\subsection{Recursive equations}
\begin{figure}[h]	
	\centering
	{\includegraphics[width=0.50\textwidth]{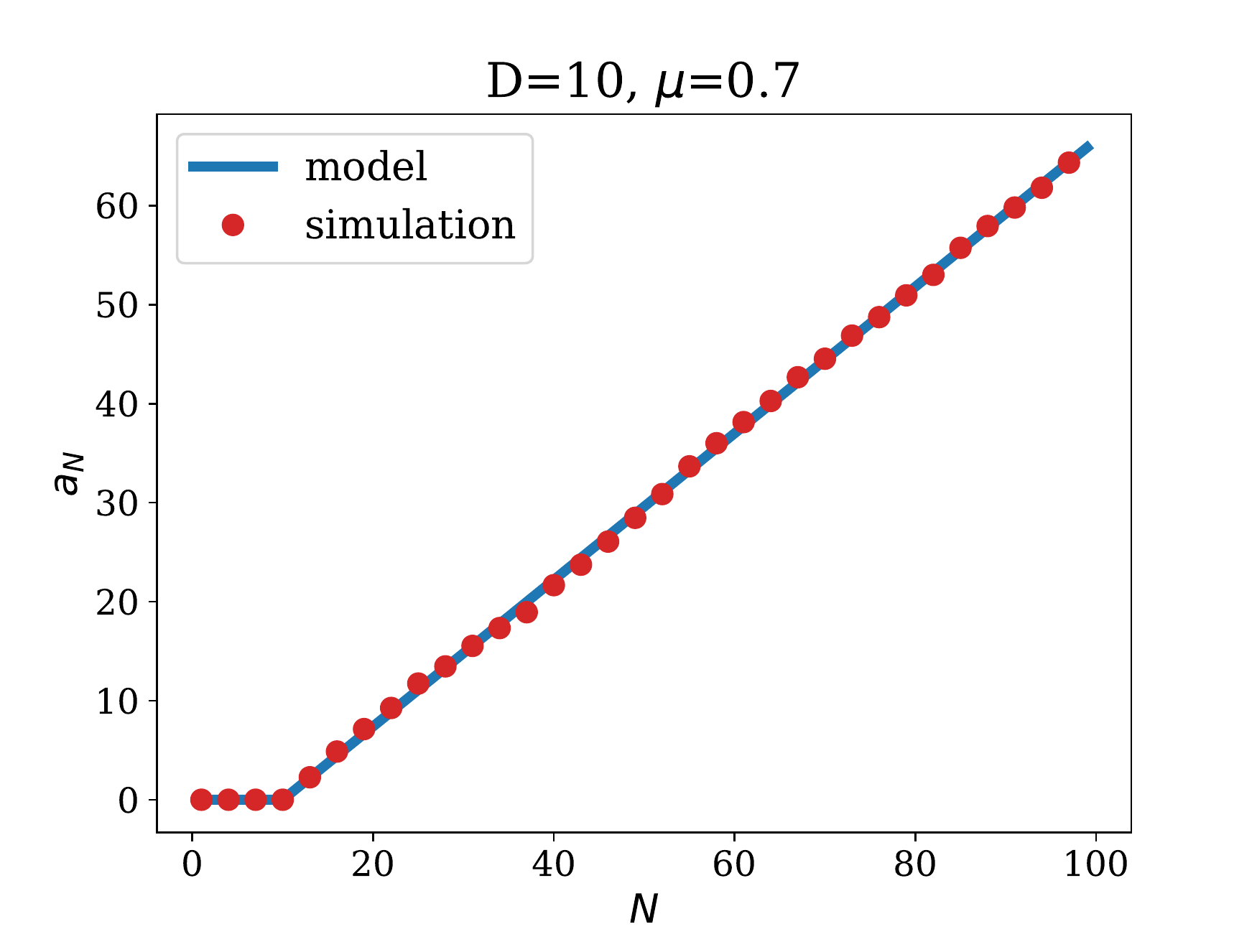}}
	\caption{Plots of $a_N$ as a function of the network size $N$, with parameters $\mu=0.7$ and $D=10$. The simulation data were obtained averaging over 20 realizations of the network generative process.}
	\label{fig:a_N}
\end{figure}

We now construct a recursion for the term $a_N$. We note that the term $a_{N+1}$ can be split into the contribution from the new row/column and the rest of the matrix as follows
\begin{eqnarray}
a_{N+1} &=& a_N + 2 \left<\sum_{r:\sigma_r=\sigma_{N+1}}^N A_{r,N+1}\right>\ .
\label{eq:aN_recursion_generic}
\end{eqnarray}
We recognize that the expectation value represents the probability that node $N+1$ attaches to any node within its own community at step $N$. Under the assumption that any target constituency is chosen independently of the degrees of its members, and using Eq. \eqref{eq:prob_si_sN}, one can derive the expression provided in \eqref{eq:a_N_seq}, as shown in more detail in the Appendix
\begin{align}
\left<\sum_{r:\sigma_r=\sigma_{N+1}}^N A_{r,N+1}\right> =
\begin{dcases}
0  &\textrm{when }N< D\\ 
1 - \mu \frac{D-1}{D} & \textrm{when }N\geq D\ ,
\end{dcases}
\label{eq:a_N_seq}
\end{align}
where the sum runs over all pre-existing nodes (up to $N$) belonging to the community $\sigma_{N+1}$. 
Using Eq. \eqref{eq:a_N_seq}, the solution of the recursion \eqref{eq:aN_recursion_generic} is found as
\begin{equation}
a_{N} = 
\begin{dcases}
0 \quad & \textrm{when }N\leq D\\
2(N-D) \left(1 - \mu \frac{D-1}{D}\right) \quad& \textrm{when }N>D \ .
\end{dcases}
\label{eq:a_N_sol}
\end{equation}
A comparison between Eq. \eqref{eq:a_N_sol} and a numerical simulation is shown in Figure \ref{fig:a_N}.

We then consider the recursion for term $b_N$. Following our definition in Eq. \eqref{eq:modularity_Newm_def}, 
 \begin{align}
 b_{N+1} = \underbracket[0.8pt]{ \left<\sum_{s,r}^{N} k_r(N+1)k_s(N+1)\delta_{\sigma_r,\sigma_s}\right> }_{(i)}\nonumber\\ +2 \underbracket[0.8pt]{\left<\sum_{r}^{N} k_r(N+1)\delta_{\sigma_r,\sigma_{N+1}}\right>}_{(ii)}+1 \ ,
\label{eq:b_N_seq}
\end{align}
where we used $k_{N+1}(N+1)=m=1$. Distinguishing the two cases
\begin{itemize}
    \item \textbf{case $\mathbf{N\geq D}$}: the expectation $(i)$ in Eq. \eqref{eq:b_N_seq} yields 
\begin{align}
&\left<\sum_{s,r}^{N} k_r(N+1)k_s(N+1)\delta_{\sigma_r,\sigma_s}\right>=\nonumber\\
&\quad= b_N+ 4\frac{\mu}{D}(N-1) + 2\left(1-\mu\right)C_{N+1}(N)+1\ ,
\end{align}
as discussed in the Appendix, and for the expectation $(ii)$ in Eq. \eqref{eq:b_N_seq}  
\begin{align}
    	\left<\sum_{r:\sigma_r=\sigma_{N+1}}^{N} k_r(N+1) \right> = C_{N+1}(N) + 1-\mu\frac{D-1}{D}\ ,
\end{align}
also in the Appendix, where
\begin{align}
&C_{N+1}(N) = \left<\sum_{r:\sigma_r=\sigma_{N+1}} k_r(N)\right> 
\label{eq:C_N+1_N_def}
\end{align}
represents the average number of intra-cluster connections for constituency $\sigma_{N+1}$ at time $N$. When evaluating the expectation in Eq. \eqref{eq:C_N+1_N_def} one gets
\begin{align}
&C_{N+1}(N) = \left[ \sum_{x=\textrm{mod}(N+1,D)+1}^D \frac{1}{x-1} + \frac{\mu}{D}\times\right.\nonumber\\
&\times(\textrm{mod}(N+1,D)-1) \Bigg](1-\delta_{\textrm{mod}(N+1,D),0}) +\nonumber\\
&+\mu\frac{D-1}{D}\delta_{\textrm{mod}(N+1,D),0} + 2 \left\lfloor\frac{N}{D}\right\rfloor -1 \ ,
\label{eq:L_N+1_N}
\end{align}
where $\textrm{mod}(\ \cdot\ ,D)$ is the modulus operator with divisor $D$ and $\left\lfloor\cdot\right\rfloor$ denotes the floor operation.

\item \textbf{case $\mathbf{N<D}$}: this case is characterized by only $N$ constituencies being yet populated and a uniform probability of wiring, leading to the following expectation for $(i)$ in Eq. \eqref{eq:b_N_seq} 
\begin{align}
&\left<\sum_{s,r}^{N} k_r(N+1)k_s(N+1)\delta_{\sigma_r,\sigma_s}\right>=\nonumber\\
&\quad = b_N+ 4\frac{N-1}{N} +1\ ,
\end{align}
as shown in the Appendix, while the expectation $(ii)$ in Eq. \eqref{eq:b_N_seq} reads
\begin{align}
    	\left<\sum_{r:\sigma_r=\sigma_{N+1}}^{N} k_r(N+1) \right> = 0\ ,
\end{align}
since constituency $\sigma_{N+1}$ is populated by one node at time step $N+1$, which is however excluded from the sum.
\end{itemize}

\begin{figure}[h]
	\centering
{\includegraphics[width=0.50\textwidth]{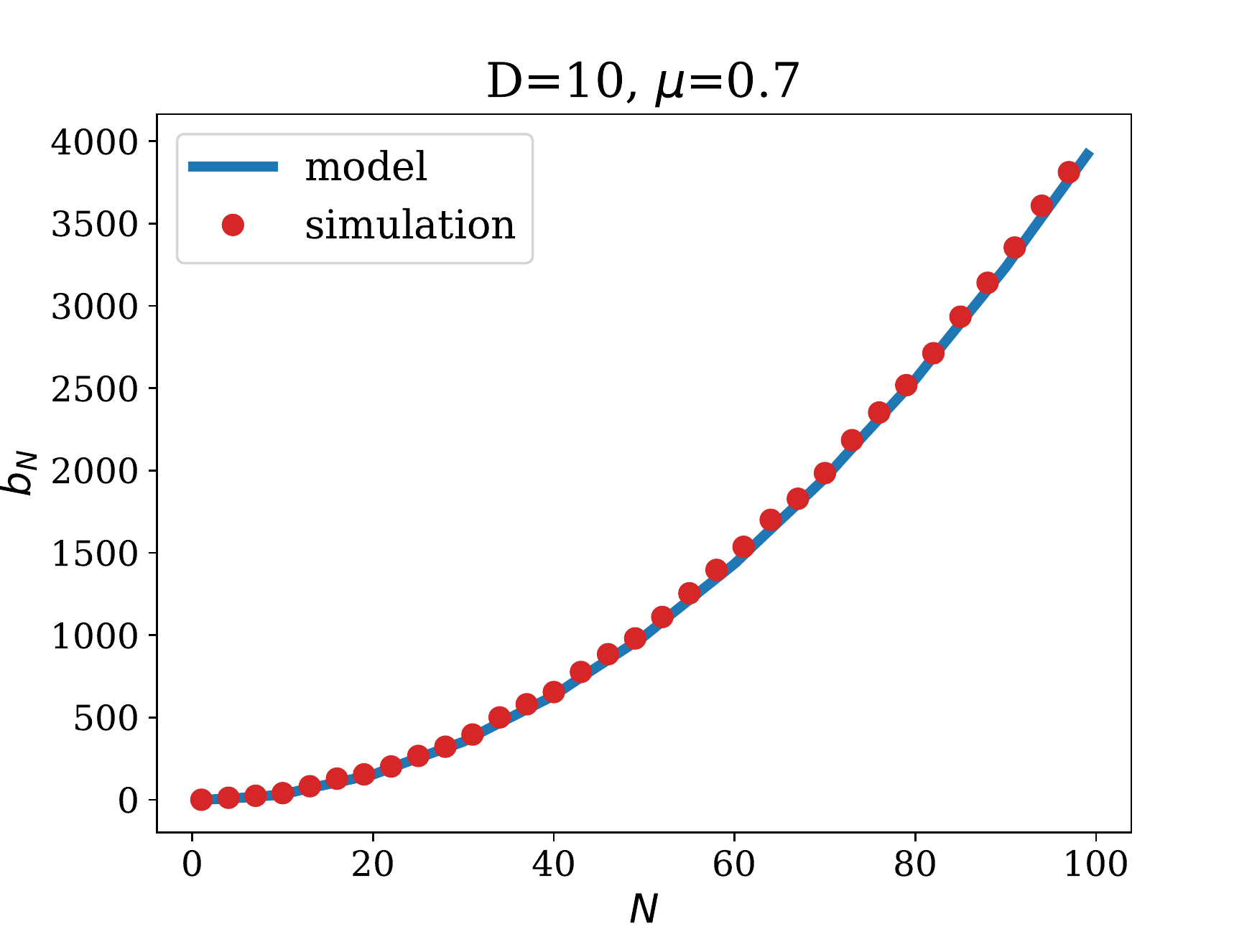}}
	\caption{$b_N$ as a function of the network size $N$ for the parameters $D=10$ and $\mu=0.7$. The simulation data were obtained averaging over 20 realizations of the network generative process.}
	\label{fig:b_N_full}
\end{figure}
Gathering all the terms, the recursive equation for $b_N$ is found to be
\begin{widetext}
\begin{align}
b_{N+1} = 
\begin{dcases}
b_{N} + 4 \frac{N-1}{N} +2 & \textrm{when } 2< N<D\\ 
b_{N} + 4 \mu\frac{N-1}{D} +2(2-\mu) C_{N+1}(N)+2\left(2-\mu\frac{D-1}{D}\right) & \textrm{when }N\geq D\ ,
\end{dcases}
\label{eq:b_N_ND}
\end{align}
\end{widetext}
with initial condition 
\begin{eqnarray}
b_2 = 2 +2\delta_{D,1} \ .
\label{eq:init_cond_b_n}
\end{eqnarray}

Solving the recursion in Eq. \eqref{eq:b_N_ND}, for general $N$, is not an easy task. In the next Section, we provide an analytical expression for  $C_{N+1}(N)$, in the limit $N\gg D$, which turns out to be a good approximation for the exact solution, even at small $N$. In Figure \ref{fig:b_N_full} we plot the numerical solution for $b_N$ against numerical simulations of the growing process.

\subsection{Approximate solution}
\begin{figure}[h]
	\centering
	{\includegraphics[width=0.50\textwidth]{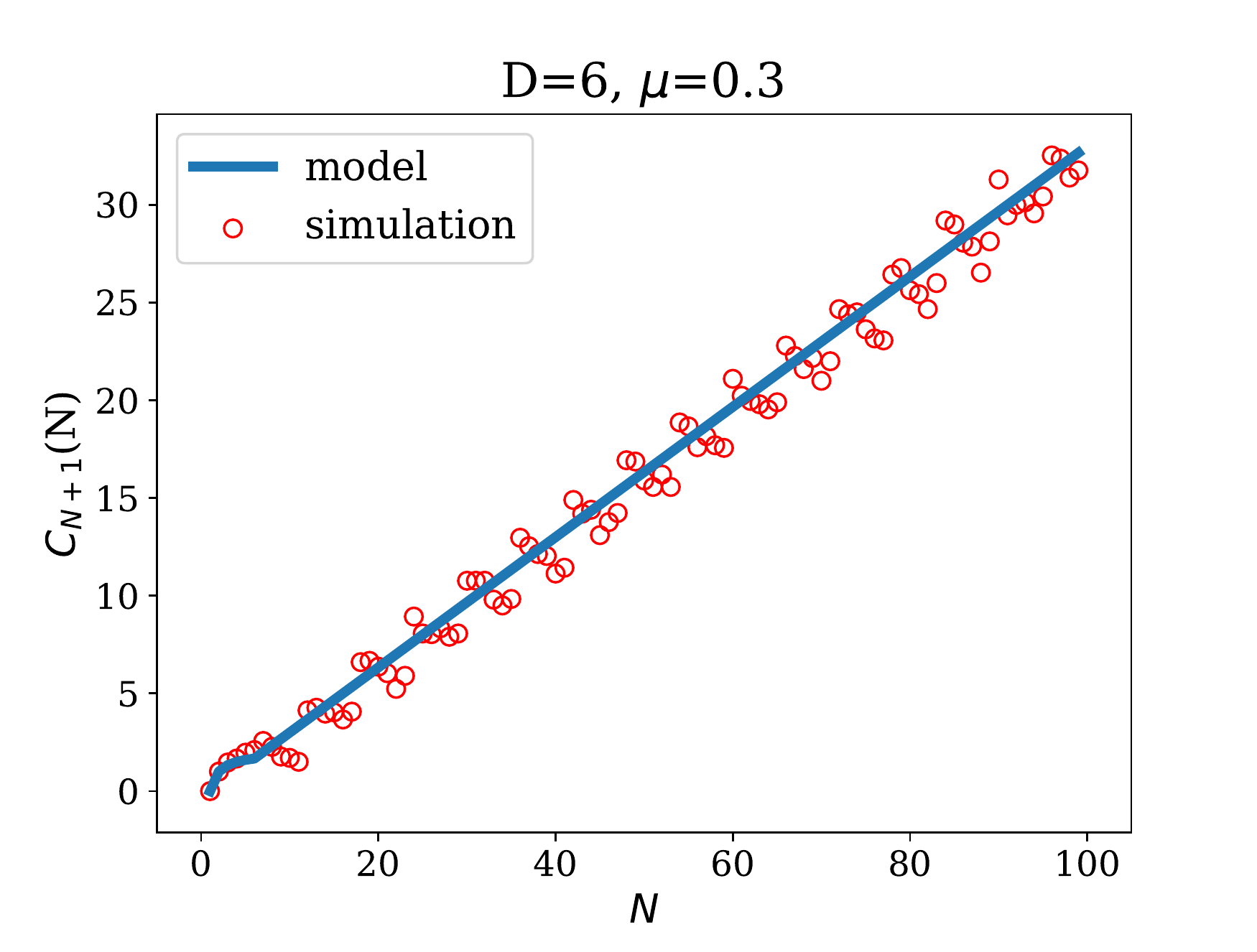}}
	\caption{Linear approximation of $C_{N+1}(N)$ plotted against numerical simulations for $D=6$ and $\mu=0.3$. The simulation data were obtained averaging over 20 realizations of the same experiment.}
	\label{fig:L_N}
\end{figure}

To make analytical progress, we assume that the edges are uniformly distributed between the $D$ communities so that $C_{N+1}(N)\simeq L(N)/D$. This holds true in the limit $N\gg D$, however, uniformity is not expected in the regime $N\sim D$, as shown in Figure \ref{fig:L_N}. The exact calculation for the expression in Eq. \eqref{eq:L_sig} in the regime $N<D$ is carried out in the Appendix. Combining the results in the two regimes, we have
\begin{equation}
\left<\sum_{s,r}^{N} k_r(N)A_{N+1 s} \delta_{\sigma_r,\sigma_s} \right>= 
\begin{dcases}
2\frac{(N-1)}{N}  & \textrm{when }N< D\\
2\frac{(N-1)}{D} & \textrm{when }N\gg D\ . 
\end{dcases}
\label{eq:L_sig}
\end{equation}
The expression provided for $N\gg D$ turns out however to accurately capture the trend for in Eq. \eqref{eq:L_sig}, even for finite $N\geq D$, and thus we will extend it to the whole range of $N$. 

\begin{figure}[h]
	\centering
	{\includegraphics[width=0.50\textwidth]{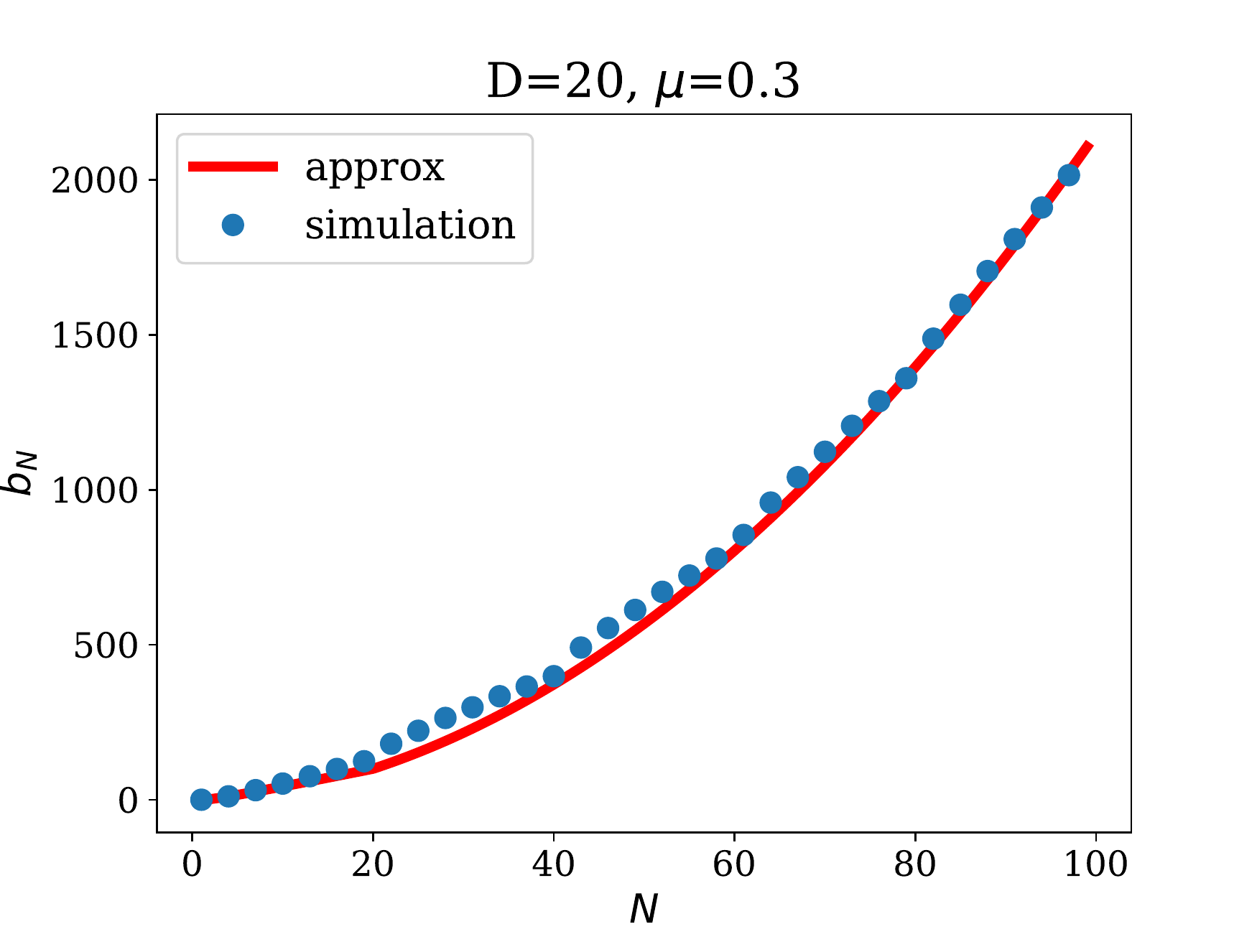}}
	\caption{Approximate solution for $b_N$ plotted against numerical simulations of the generative process for $D=20$ and $\mu=0.3$. The simulation data were obtained averaging over 50 realizations of the same experiment.}
	\label{fig:b_N}
\end{figure}

Under this approximation, the recursion in Eq. \eqref{eq:b_N_seq}, now simplifies to
\begin{align}
	b_{N+1} =
	\begin{dcases}
	 b_{N} + 4 \frac{N-1}{N} +2 & \textrm{when } 2< N<D\\
	 b_{N} + 4 \frac{2N-1}{D} +2  & \textrm{when }N\geq D\ ,
	\end{dcases}
\end{align}
with the initial condition \eqref{eq:init_cond_b_n}.
The solution is given by
\begin{widetext}
\begin{align}
b_N =
\begin{dcases}
-2(3+2\gamma -3N+2\psi^{(0)}(N)) &\textrm{when }N< D\\
\frac{2}{D}( D -4N +DN + 2N^2 - 2D(\gamma+\psi^{(0)}(D)) &\textrm{when }N\geq D \ ,
\end{dcases}
\label{eq:b_N_sol}
\end{align}
\end{widetext}
with $\gamma$ being the Euler-Mascheroni constant and $\psi^{(0)}(x)$ the digamma function, arising from summing the first inverse integers series $\sum_{k=1}^{x-1}\frac{1}{k}$ .
This approximate solution is in very good agreement with numerical simulations, as shown in Figure \ref{fig:b_N}.

Inserting in Eq. \eqref{eq:modularity_Newm_def} the expressions for $a_N$ and $b_N$, provided in Eq. \eqref{eq:a_N_sol}  and in Eq. \eqref{eq:b_N_sol} respectively, we obtain the solution for $\left<Q_D(N)\right>$ in the regime $N\geq D$ 
\begin{align}
\left<Q_D(N)\right> =
\frac{N-D}{N-1} \left(1 - \mu \frac{D-1}{D}\right) -\frac{1}{2D(N-1)^2}\times\nonumber\\
\times (D -4N +DN+ 2N^2- 2D(\gamma+\psi^{(0)}(D))) \ . 
\label{eq:Q_recurrence}
\end{align}

\begin{figure}[h]
	\centering
	{\includegraphics[width=0.50\textwidth]{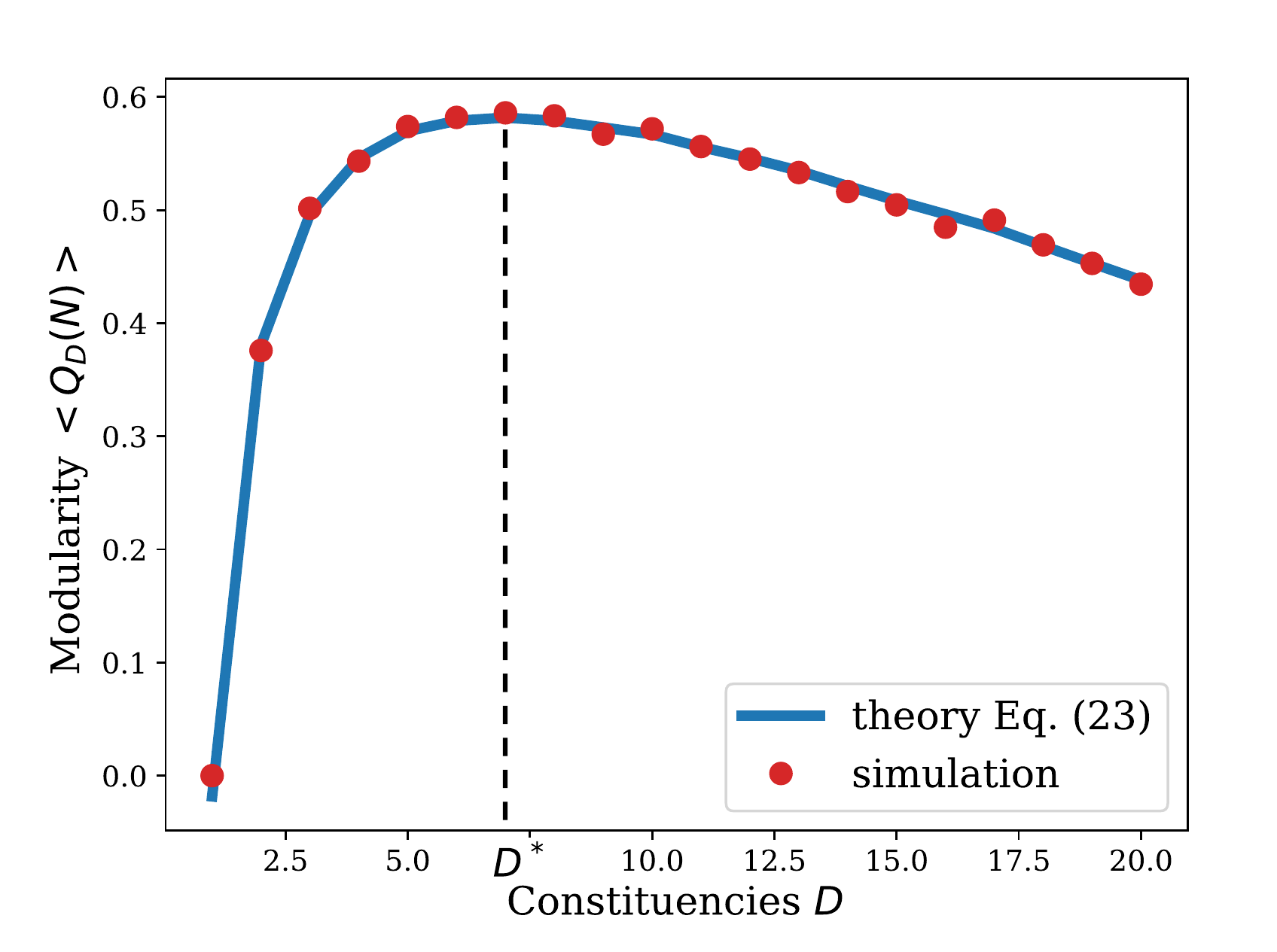}}
	\caption{Expression for the average modularity $\left<Q_D(N)\right>$ in Eq. \eqref{eq:Q_recurrence} as a function of the number of constituencies $D$, with parameters $\mu=0.2$ and $N=50$. The simulation data were obtained averaging over 30 realizations of the same experiment.}
	\label{fig:Q_D}
\end{figure}

The numerical simulation in Figure \ref{fig:Q_D} shows a perfect agreement with the average modularity given by Eq. \eqref{eq:Q_recurrence}. 
We observe that the expected modularity is strongly non-monotonic in the number of constituencies. This is in agreement with the following observations about the limiting cases: for $D=1$, the two terms of the sum in Eq. \eqref{eq:modularity_Newm_def_non_exp} cancel out, and for $D=N$ the concept of community is lost and the Kronecker delta is always zero, thus both cases result in $Q_D(N)=0$. Note that in the intermediate regime $1<D<N$ and provided that $\mu\in[0,1)$ we have that the probability for a node to link to any of its fellow constituents is higher than the ``rest of the population". This ensures that, on average, the modularity is positive. This non-monotonic behavior was also observed empirically in \cite{HBB09}.

Values for $\mu\in[0,1)$ are consistent with the constituencies design process according to which boundaries should be drawn around local communities. The regime $\mu\geq 1$ would result in an equal or lower intra-constituency edge density compared to the density of outgoing edges, suggesting that the imposed partitions would not capture the real community structure of the network and thus would not be interesting for our purpose. Moreover, $\mu=1$ is the physiological upper bound to ensure that probability \eqref{eq:prob_si_sN} is non-negative.

\begin{figure}[h]
	\centering
	{\includegraphics[width=0.50\textwidth]{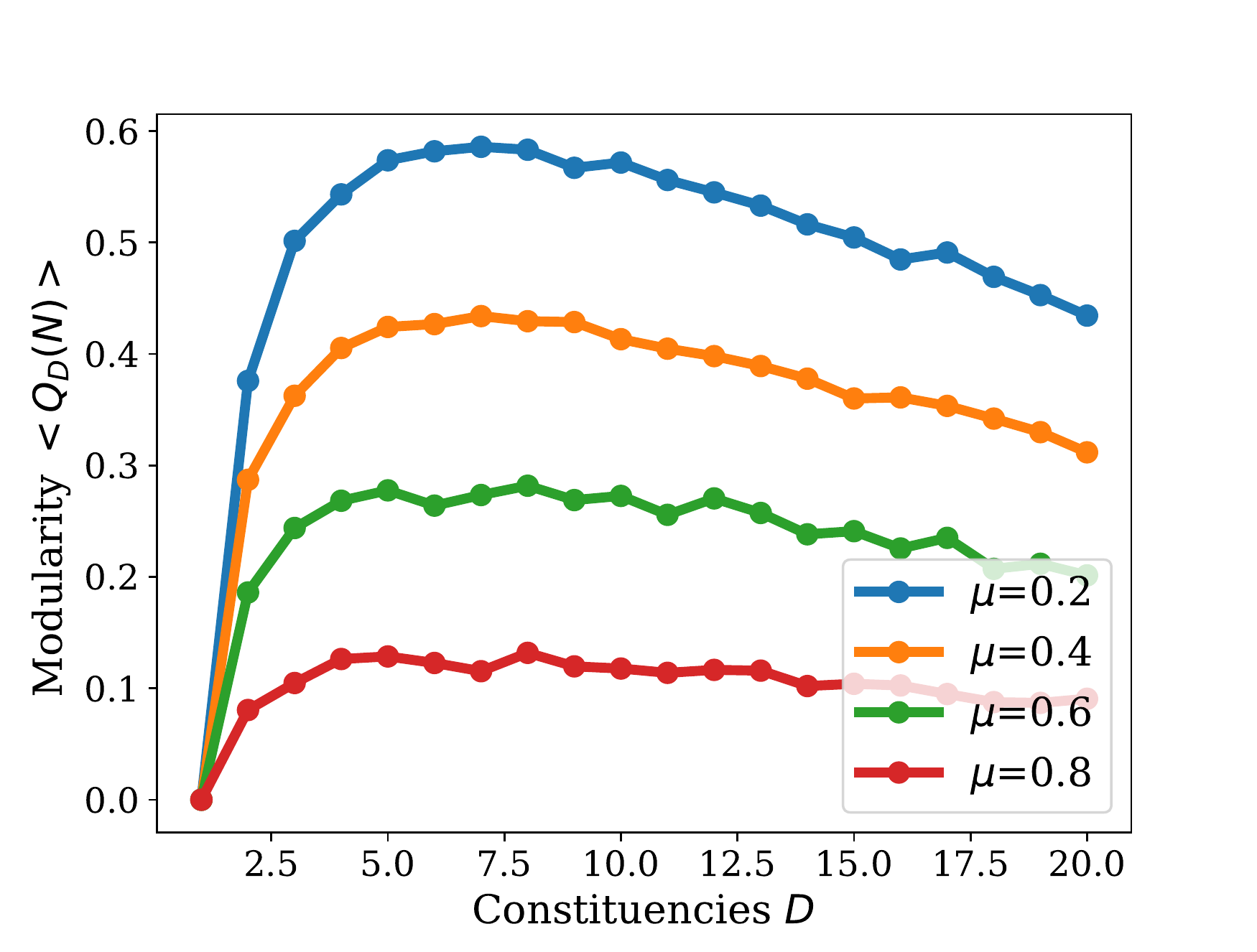}}
	\caption{The plot shows the effect of the mobility parameter $\mu$ on the modularity $\left<Q_D(N)\right>$, with $N=50$. The simulation data were obtained averaging over 20 realizations of the same experiment.}
	\label{fig:Q_D_vs_mu}
\end{figure}

Furthermore, we observe that the mobility parameter $\mu$ dampens the modularity without producing a pronounced shift of its maximum, as shown in Figure \ref{fig:Q_D_vs_mu}. This effect is due to a tightening of the community structures within each constituency as the effect of a decreasing $\mu$ is to increase their average intra-cluster density and thus to increase the overall modularity. Conversely, when $\mu \to 1$, nodes attach randomly to any constituency resulting in an average modularity $Q_D(N)$ that tends to zero.

\begin{figure}[h]
	\centering
	{\includegraphics[width=0.5\textwidth]{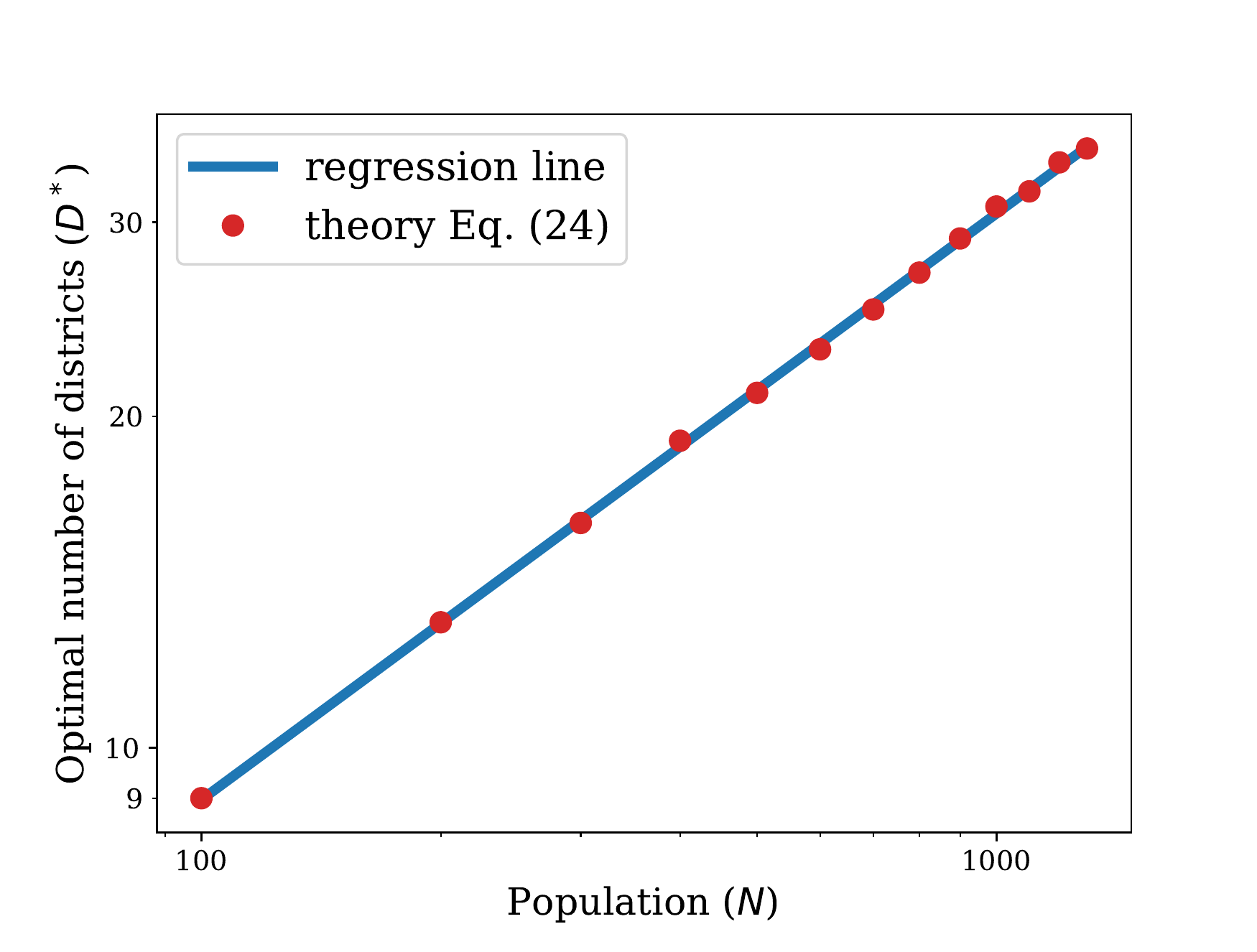}}
	\caption{The log-log plot shows the power-law fit for the position of the maximum modularity $D^* = \alpha N^\gamma$, in the linear approximation regime, for $\mu=0.9$, resulting to an $\alpha\approx 0.77$ and $\gamma\approx0.53$.}
	\label{fig:D*_N}
\end{figure}

Finally, we find an expression for the maximum value of the modularity. Indeed, it is our key objective to find an optimal way to partition our synthetic population into constituencies. We argue that this optimal way of partitioning is realized when the modularity reaches its maximum and the imposed partitions best capture the underlying community structure of the network. We derive then the location $D^*(N)=\arg\max_D \left<Q_D(N)\right>$, in the regime $N\gg D$, from Eq. \eqref{eq:Q_recurrence}
\begin{align}
&\frac{\partial \left<Q_D(N)\right>}{\partial D} = \frac{1}{(N-1)^2}[ (N-1)(\mu -1) +\nonumber\\
&\quad + \frac{1}{D^{*2}}N(N+\mu(1-N) -2) + \psi^{(1)}(D^*) ]=0 \ ,
\label{eq:mod_derivative}
\end{align}
with $\psi^{(1)}(D)$ being the first order polygamma function, defined as the first derivative of the digamma function.
This expression constitutes our main result, as it gives a recipe to pick the optimal number of constituencies, for a given population size $N$. The implicit Eq. \eqref{eq:mod_derivative} can be solved numerically for $D^*$. Interestingly, we find that $D^*(N)$ has a clear power-law behavior similar to the one observed in demographic data. 
Figure \ref{fig:D*_N} shows $D^*(N)$ for small networks, the numerical data being perfectly fitted by power-law $D^*=\alpha N^\gamma$, with exponent $\gamma \approx 0.53$ and $\alpha\approx 0.77$.
The value of the exponent can be also determined by the following analytical consideration. Expression \eqref{eq:mod_derivative} can be rewritten as follows
\begin{equation}
	\alpha_N + \frac{1}{D^{*2}}\beta_N + \psi^{(1)}(D^*) = 0\ ,
	\label{eq:D*_analit_arg}
\end{equation} 
with the coefficients $\alpha_N = (N-1)(\mu -1)$ and $\beta_N = N(N+\mu(1-N)-2)$. By setting $\psi^{(1)}(D^*) = T$ and extracting $D^*$ from Eq. \eqref{eq:D*_analit_arg} 
\begin{equation}
	T = \psi^{(1)}\left(\sqrt{\frac{-\beta_N}{\alpha_N + T}}\right) \ .
	\label{eq:T_polygamma}
\end{equation}
Now, since we are evaluating this quantities in the large network limit, we may use the polygamma asymptotic behaviour $\psi^{(1)}(x)\sim \frac{1}{x}$ in Eq. \eqref{eq:T_polygamma}, and solve for $T$ obtaining
\begin{equation}
T \approx \frac{1}{2}\left( -\frac{1}{\beta_N} + \frac{\sqrt{1-4\alpha_N\beta_N}}{\beta_N}\right)\ .
\label{eq:T_approx}
\end{equation}
Inserting Eq. \eqref{eq:T_approx} in the original Eq. \eqref{eq:D*_analit_arg}, we obtain the asymptotic optimal number of constituencies 
\begin{equation}
	D^* \approx \sqrt{\frac{-\beta_N}{\alpha_N + \frac{1}{2}\left( -\frac{1}{\beta_N} + \frac{\sqrt{1-4\alpha_N\beta_N}}{\beta_N}\right)}} \sim \sqrt{N} \ ,
\end{equation}
which is consistent with the numerical solution of the implicit Eq. \eqref{eq:mod_derivative} shown in Figure \ref{fig:D*_N}.

\section{Conclusion}
The problem of democratic representation is of primary importance for modern societies. In this work, we proposed a network model representing a growing population of final size $N$ that has to be partitioned into $D$ equally sized constituencies. The underlying network community structure can be tuned by the mobility parameter $\mu$ that controls the interaction probability between nodes belonging to different constituencies. 

We adopted the average modularity as a measure for the goodness of the resulting partitioning and shown that it displayed a strong non-monotonic behavior, as a function of $D$, in the regime $\mu\in[0,1)$.  By solving the recurrence equations for the modularity in the regime $N\gg D$, we found an analytical expression for the optimal number of constituencies $D^*$ that maximizes the modularity w.r.t. the number of induced partitions.

The approximate regime in which the problem is solved corresponds to one MP accounting for a large fraction of the population. This is arguably a reasonable assumption when considering democratically elected parliaments for which the condition above is always satisfied. Nevertheless, a numerical solution was also attainable for any value of $N$ and $D$.
Our main finding concerns the functional form for the optimal size of a Parliament $D^*\sim N^{1/2}$ that is found to be in reasonable agreement to what is observed in real-world data, for European parliaments see Figure \ref{fig:euparl_vs_pop}.

While a larger mobility parameter induces a more efficient mixing of the population and therefore reduces its average modularity for a fixed number of available constituencies (see Fig. \ref{fig:Q_D_vs_mu}), quite interestingly it does not influence the position of the maximum $D^*$ as a function of $N$ to leading order. 
This is due to the assumed homogeneity of the network, where the mobility
is fixed to a constant value for the entire population. We suggest that as pathways for future work one could consider introducing geographical constraints in the model and including a mobility parameter that depends on the population density of each constituency. This is expected to generate a richer behavior for $D^*$. 

\subsection*{Acknowledgements} 
PV and ET acknowledge support from UKRI Future Leaders Fellowship scheme n. MR/S03174X/1. Also, support by EPSRC through EP/L015854/1 Centre for Doctoral Training CANES (from Y-PF and AA) is thankfully
acknowledged.
\newpage

\newpage
\onecolumngrid
\vspace{\columnsep}

\section*{\LARGE Appendix}
In this section, we discuss the recursion for $a_N$ and we present a more detailed version of the exact calculation for $b_N$. 

\paragraph*{$\bm{a_N}$:}We consider the term $\left<\sum_{r:\sigma_r=\sigma_{N+1}}^N A_{r,N+1}\right>$, appearing in Eq. \eqref{eq:aN_recursion_generic}. The expectation over the distribution of Eq. \eqref{eq:mobility_ensemble} is given by
\begin{align}
   \left<\sum_{r:\sigma_r=\sigma_{N+1}}^N A_{r,N+1}\right>_{p(\mathbf{A}(N+1)|\boldsymbol{\sigma}(N+1))}&=\left<\sum_{r:\sigma_r=\sigma_{N+1}}^N \sum_{A_{r,N+1}=\{0,1\}} A_{r,N+1} p(A_{r,N+1}|\mathbf{A}(N),\boldsymbol{\sigma}(N))\right>_{p(\mathbf{A}(N)|\boldsymbol{\sigma}(N))}\nonumber\\
   &=\left<\sum_{r=1}^N \delta_{\sigma_r,\sigma_{N+1}} \frac{k_r(N)}{L(N)/D} p(\sigma_{r}|\sigma_{N+1})\right>_{p(\mathbf{A}(N)|\boldsymbol{\sigma}(N))}\nonumber\\
   &= p(\sigma_{N+1}|\sigma_{N+1}) \frac{1}{L(N)/D}\left<\sum_{r=1}^N k_r(N) \delta_{\sigma_r,\sigma_{N+1}} \right>_{p(\mathbf{A}(N)|\boldsymbol{\sigma}(N))}\label{sums_app}\ ,
\end{align}   
where we have used Eq. \eqref{eq:mobility_attachment_prob}. Note that the term within angle brackets is zero for $N<D$, as in this case no node is present in the network with the relevant label. For $N$ large, the average of the sum in \eqref{sums_app} converges to $N\sum_k ~k p(k,\sigma_{N+1})$, in terms of the joint probability $p(k,\sigma_{N+1})$. This in turn factorizes in the product of marginals due to the independence of $k$ and $\sigma$, leading eventually to the result
\begin{equation}
 \left<\sum_{r:\sigma_r=\sigma_{N+1}}^N A_{r,N+1}\right>_{p(\mathbf{A}(N+1)|\boldsymbol{\sigma}(N+1))}\simeq p(\sigma_{N+1}|\sigma_{N+1})\ ,
\end{equation}
where one uses $L(N)=N\langle k(N)\rangle$, and $p(\sigma_{N+1})=1/D$. For $N\geq D$ but not too large, fluctuations of $\mathcal{O}(1/N)$ are expected.

\paragraph*{$\bm{b_N}$:}Recall the expression from Eq. \eqref{eq:b_N_seq}
\begin{eqnarray}
 b_{N+1} &=& \underbracket[0.8pt]{ \left<\sum_{s,r}^{N} k_r(N+1)k_s(N+1)\delta_{\sigma_r,\sigma_s}\right> }_{(i)} +2 \underbracket[0.8pt]{\left<\sum_{r}^{N} k_r(N+1) \delta_{\sigma_r,\sigma_{N+1}} \right>}_{(ii)}+1 \ ,
 \end{eqnarray}
where the factor $2$ is due to the symmetry of the adjacency matrix.
We analyze each average in Eq. \eqref{eq:b_N_seq}, labelled $(i)$ and $(ii)$, separately:
\begin{itemize}
	\item[$(i)$]
Starting with the case $N\geq D$, we have observed that, when $m=1$, the degree of a given node can only increase by one at each time step and only if the newly created node connects to it, i.e.   $k_r(N+1)= k_r(N) + A_{r, N+1}$. This leads to
 \begin{align}
\left<\sum_{s,r}^{N} k_r(N+1)k_s(N+1)\delta_{\sigma_r,\sigma_s}\right>&=\left<\sum_{s,r}^{N} (k_r(N)+A_{r,N+1})(k_s(N)+A_{s,N+1})\delta_{\sigma_r,\sigma_s}\right>\nonumber\\
&= \left<\sum_{s,r}^{N} k_r(N)k_s(N)\delta_{\sigma_r,\sigma_s}\right> + 2\left<\sum_{s,r}^{N} k_r(N)A_{N+1, s} \delta_{\sigma_r,\sigma_s} \right> +\nonumber\\
&\quad +\left<\sum_{s,r}^{N}  A_{r, N+1}A_{N+1, s} \delta_{\sigma_r,\sigma_s} \right>\nonumber\\
&= b_N + 2\left<\sum_{s,r}^{N} k_r(N)A_{N+1, s} \delta_{\sigma_r,\sigma_s} \right> +\left<\sum_{s,r}^{N}  A_{r, N+1}A_{N+1, s} \delta_{\sigma_r,\sigma_s} \right>\ .
\label{eq:b_N+1_i}
\end{align}
The first expectation may now be rewritten in the following way
\begin{align}
\left<\sum_{s,r}^{N} k_r(N)A_{N+1,s} \delta_{\sigma_r,\sigma_s} \right> &= \left<\sum_{s}^{N} A_{N+1, s}\sum_{r:\sigma_r = \sigma_s}^{N} k_r(N) \right>\nonumber\\
&=\left<\sum_{s:\sigma_s=\sigma_{N+1}} A_{N+1, s} \sum_{r:\sigma_r = \sigma_{N+1}} k_r(N) \right> +
\left<\sum_{s:\sigma_s\neq\sigma_{N+1}} A_{N+1, s} \sum_{r:\sigma_r \neq \sigma_{N+1}} k_r(N) \right> \ .
\label{eq:first_exp_value}
\end{align}
The $D-1$ constituencies such that $\sigma_r\neq \sigma_{N+1}$ are equivalent in the sum above as the probability of wiring $N+1\sim r$ is $\mu/D$ for all, thus, considering the contribution to Eq. \eqref{eq:first_exp_value} from one constituency $\sigma_{\bar{s}}\neq \sigma_{N+1}$ we may rewrite
\begin{align}
\left<\sum_{s,r}^{N} k_r(N)A_{N+1,s} \delta_{\sigma_r,\sigma_s} \right> &= \left<\sum_{s:\sigma_s=\sigma_{N+1}} A_{N+1, s} \sum_{r:\sigma_r = \sigma_{N+1}} k_r(N) \right> + (D-1)
\left<\sum_{s:\sigma_s=\sigma_{\bar{s}}} A_{N+1, s} \sum_{r:\sigma_r =\sigma_{\bar{s}}} k_r(N) \right> \ .
\label{eq:first_exp_value_ii}
\end{align}
Now performing the expectation value of $\sum_{s:\sigma_s=\sigma_{N+1}} A_{N+1, s}$, using Eq. \eqref{eq:a_N_seq},  we get
\begin{align}
\left<\sum_{s,r}^{N} k_r(N)A_{N+1, s} \delta_{\sigma_r,\sigma_s} \right> = \left(1 - \mu \frac{D-1}{D}\right)\left<\sum_{r:\sigma_r = \sigma_{N+1}}^{N} k_r(N) \right> + \mu\frac{D-1}{D}\left< \sum_{r:\sigma_r =\sigma_{\bar{s}}} k_r(N) \right> \ .
\label{eq:first_exp_value_2}
\end{align}
We are left with the term $\left< \sum_{r:\sigma_r=\sigma_{N+1}}^N k_r(N)\right> =C_{N+1}(N)$ to calculate, being the expected number of links in community $\sigma_{N+1}$ at time $N$ and its complementary $\left< \sum_{r:\sigma_r\neq\sigma_{N+1}} k_r(N)\right> =\bar{C}_{N+1}(N)$.

When the first round of communities has been assigned, i.e. $N\geq D$, the expression becomes 
\begin{eqnarray}
\left<\sum_{s,r}^{N} k_r(N)A_{N+1 s} \delta_{\sigma_r,\sigma_s} \right> &=&\left(1-\mu\frac{D-1}{D}\right)C_{N+1}(N) +\mu\frac{D-1}{D}\frac{\bar{C}_{N+1}(N)}{D-1}\nonumber\\
&=&\left(1-\mu\frac{D-1}{D}\right)C_{N+1}(N) +\frac{\mu}{D}(2(N-1)-C_{N+1}(N))\nonumber\\
&=&2\frac{\mu}{D}(N-1) + \left(1-\mu\right)C_{N+1}(N) \ ,
\label{eq:L_sig_sol}
\end{eqnarray}
where we have used $\bar{C}_{N+1}(N) + C_{N+1}(N) = 2(N-1)$. 
The expression for $C_{N+1}(N)$ is found to be 
\begin{eqnarray}
C_{N+1}(N) &=& \left[ \sum_{x=\textrm{mod}(N+1,D)+1}^D \frac{1}{x-1} + \frac{\mu}{D}(\textrm{mod}(N+1,D)-1) \right](1-\delta_{\textrm{mod}(N+1,D),0}) +\nonumber\\
& &+\mu\frac{D-1}{D}\delta_{\textrm{mod}(N+1,D),0} + 2 \left\lfloor\frac{N}{D}\right\rfloor -1 \ ,
\label{eq:L_N+1}
\end{eqnarray}
where $\textrm{mod}(\ \cdot\ ,D)$ is the modulus operator with divisor $D$ and $\left\lfloor\cdot\right\rfloor$ denotes the floor operator. This builds up by summing the contribution to the expected links in community $\sigma_{N+1}$ from every node. For the case $N\geq D$, a node $j$ such that its membership  $\sigma_{j}=\sigma_{N+1}$ always contributes with one link (its own stub) plus another link with probability $1-\mu \frac{D-1}{D}$, in case the target node at the other end of the stub has the same membership $\sigma_{N+1}$. Conversely, if $\sigma_{j}\neq\sigma_{N+1}$, the new node $j$ can only add one link to constituency $\sigma_{N+1}$ with probability $\mu/D$. In the first round of membership assignment, a contribution to the number of links can come from nodes generated following the first node in $\sigma_{N+1}$, with a probability $1/j$.
The expression can be intuitively checked following Figure \ref{fig:L_N_illustration}.
\begin{figure}[h]	
	\centering
	\includegraphics[width=0.8\textwidth]{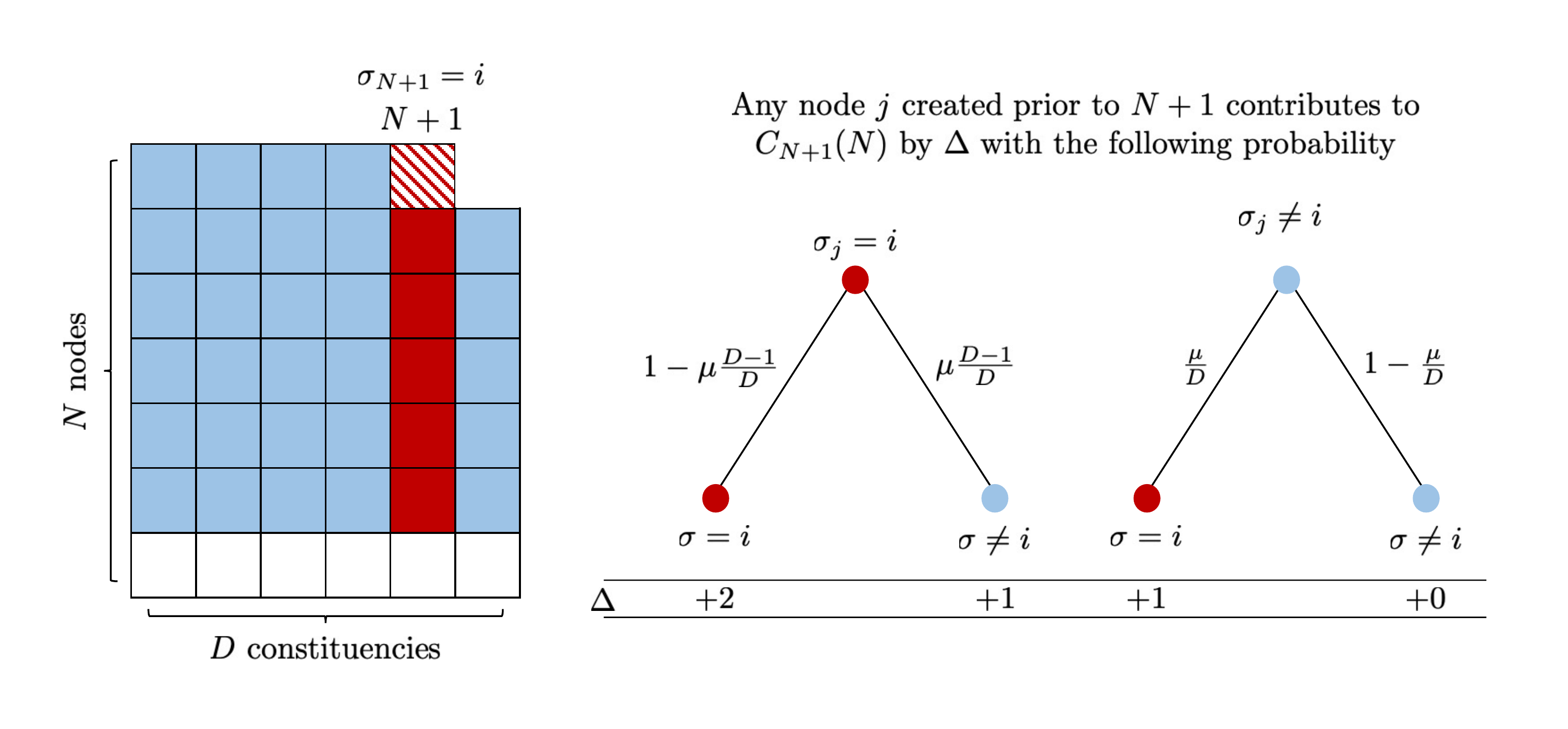}
	\caption{Illustration showing the possible increment for the number of links $C_{N+1}(N)$ in community $\sigma_{N+1}=i$ at time $N$ with the corresponding probability, for all nodes $j\geq D$ in case of sequential assignment of constituency memberships.}
	\label{fig:L_N_illustration}
\end{figure}

The remaining expectation of Eq. \eqref{eq:b_N+1_i}, is non-zero only if $s=r$, so
\begin{equation}
\left<\sum_{s,r}^{N}  A_{r, N+1}A_{N+1, s} \delta_{\sigma_r,\sigma_s} \right> = \left<\sum_{s}^{N}  A_{N+1,s} \right> = 1\ .
\end{equation}

We finally note that, in the case $N<D$, the expectations in Eq. \eqref{eq:first_exp_value} $ \left<\sum_{s:\sigma_s=\sigma_{N+1}} A_{N+1, s} \right> = 0$ and $ \left<\sum_{s:\sigma_s\neq\sigma_{N+1}} A_{N+1, s} \right> = 1$. This is due to constituency $\sigma_{N+1}$ not being populated before time $N+1$. As a result the average number of links in any given community is given by
\begin{align}
\left<\sum_{s,r}^{N} k_r(N)A_{N+1, s} \delta_{\sigma_r,\sigma_s} \right> &= \left< \sum_{r:\sigma_r \neq \sigma_{N+1}} k_r(N) \right> \nonumber\\
&=\frac{1}{N}\left< \sum_{r} k_r(N) \right> \nonumber\\
&=2\frac{N-1}{N}\ .
\label{eq:L_sig_sol_N<D}
\end{align}
	\item[$(ii)$]
Now, this expectation value is similar to $C_{N+1}(N) =\left< \sum_{r:\sigma_r=\sigma_{N+1}}^N k_r(N)\right>$. The difference lies in the time step at which this is being calculated. Thus it can be given a similar interpretation as the expected number of links in community $N+1$, at time step $N+1$, excluding the last node. This implies that if at time $N+1$ the newly created node connects to one of its own community, the quantity in question increases by one, in a similar way to what is shown in Figure \ref{fig:L_N_illustration}. We note that, since the last node is excluded, its degree shall not be counted. So, the relation with $C_{N+1}(N)$ can be made explicit as follows
	\begin{eqnarray}
		\left<\sum_{r:\sigma_r=\sigma_{N+1}}^{N} k_r(N+1) \right> = C_{N+1}(N) + 1-\mu\frac{D-1}{D}\ .
		\label{eq:L_N+1_N+1}
	\end{eqnarray}
\end{itemize} 

Back to Eq. \eqref{eq:a_N_seq}, we finally get the recursive relation we are after for the coefficient $b_N$,
\begin{equation}
	b_{N+1} = b_{N} + 4 \mu\frac{N-1}{D} +2(2-\mu) C_{N+1}(N)+2\left(2-\mu\frac{D-1}{D}\right)\ .
\end{equation}

\end{document}